\documentstyle[12pt]{article}
\input epsfig.sty
\def\beq{\begin{equation}}
\def\eeq{\end{equation}}
\def\bea{\begin{eqnarray}}
\def\eea{\end{eqnarray}}

\def\NP{{\it Nucl. Phys.} }

\def\PR{{\it Phys. Rev.} }
\def\PRL{{\it Phys. Rev. Lett.} }
\def\ap{\alpha ^{\prime}}
\def\cR{{\cal R}}
\def\cD{{\cal D}}
\def\cI{{\cal I}}
\def\cB{{\cal B}}
\begin{document}
\begin{titlepage}
Revised, July 1999 \\
\begin{flushright}
HU Berlin-EP-99/27\\
\end{flushright}
\mbox{ }  \hfill hep-th/9906073
\vspace{5ex}
\Large
\begin {center}
\bf{Concavity of the $Q\bar Q$ potential in ${\cal N}=4$ super Yang-Mills gauge theory and AdS/CFT duality}
\end {center}
\large
\vspace{1ex}
\begin{center}
H. Dorn$^{\footnotesize a}$ and V.D. Pershin$^{\footnotesize b}$
\footnote{e-mail: dorn@physik.hu-berlin.de~~~pershin@ic.tsu.ru}
\end{center}
\normalsize
\it
\vspace{1ex}
\begin{center}
$^{\footnotesize a}$
Humboldt--Universit\"at zu Berlin, Institut f\"ur Physik\\
Invalidenstra\ss e 110, D-10115 Berlin\\[2mm]
$^{\footnotesize b}$
Department of Theoretical Physics\\Tomsk State University\\
Tomsk 634050, Russia
\end{center}
\vspace{4ex}
\rm
\begin{center}
{\bf Abstract}
\end{center}
We derive a generalised concavity condition for potentials between static
sources obtained from Wilson loops coupling both to gauge bosons and a set of scalar fields. It involves the second derivatives with respect to the distance in ordinary space as well as with respect to the relative orientation in internal space. In addition we discuss the use of this field theoretical condition
as a nontrivial consistency check of the AdS/CFT duality.
\vfill      
\end{titlepage}
\section{Introduction}
The AdS/CFT duality conjecture \cite{malda0,klebanov,witten} has passed
an impressive number of consistency checks \cite{review}. However, among these
tests only few are not relying in one or another way on structures  enforced by supersymmetry and/or conformal invariance. In this situation it appears worthwhile to further analyse any possible constraint set by the first principles of
quantum field theory and to check, whether they are fulfilled by the corresponding dual partners in string theory/supergravity.

In this sense the present letter is devoted to the concavity of the potential 
between static sources in a gauge theory. In the Euclidean formulation Osterwalder-Schrader reflection
positivity \cite{os} ensures this property for potentials derived from Wilson
loops \cite{seiler,bachas}. In the AdS/CFT context the issue of concavity has been raised in ref.\cite{olesen}. But the discussion so far has not taken into account the degree of freedom connected with the relative orientation of the static sources ($Q\bar Q$) in internal space.

We will fill this gap by analysing in some detail the consequences of OS reflection positivity for potentials derived in standard manner from Wilson loops
for contours coupling both to the gauge bosons and to a set of scalar
fields in the adjoint representation.  We take the Wilson loop in the form
suggested in \cite{rey,malda} and analysed  in various ways in \cite{gross}.
For the case where the gauge bosons and the scalars are just the bosonic
fields of $D=4$, ${\cal N}=4$ super-Yang-Mills theory it has been characterised
as an object of BPS type \cite{gross}.

Our discussion closely follows \cite{bachas}. The new input in our
presentation is the handling of the contour parameter dependent coupling
to the scalars, which is described by a curve on $S^5$. We also take care
of the fact that the Wilson loop of refs.\cite{malda,gross} is the trace of 
a generically non-unitary matrix.

The virtue of the arising concavity condition lies in its inequality
property. It has to be fulfilled both for the classical SUGRA approximation
and for the expressions obtained by adding successive corrections. Therefore,
a violation at any level of approximation on the superstring side
would indicate a breakdown of the corresponding duality.
\section{Generalised concavity for potentials derived from BPS Wilson loops}
We start with the functional ($A\dot x=A_{\mu}\dot x^{\mu},~\phi\theta =
\phi _j\theta ^j,~~~\mu =0,..,3,~~j=4,..,9 $)
\beq
U_{ab}[x,\theta]=\left (P\exp\int \{iA(x(s))\dot x (s)+\phi (x(s))\theta (s)
\vert \dot x\vert\}ds \right )_{ab}~.
\label{10}
\eeq
The expectation value of its trace for a closed path $x(s)$ yields the Wilson loop under investigation \cite{malda,gross}. $\theta (s)$ specifies the coupling to the scalars $\phi $
along the contour $x(s)$.

A reflection operation ${\cal R}$ is defined by
\bea
(\cR x)^1 (s)=-x^1 (s)~;~~~~(\cR x)^{\alpha}(s)=x^{\alpha}(s),~~~ \alpha\neq 1
\nonumber\\
{\cal R} U_{ab}[x,\theta ]=\overline{U_{ab}[{\cal R} x, \theta]}~.
\label{11}
\eea
In addition, it is useful to define in connection with an isometry 
$\cI \in O(6)$ 
of $S^5$ acting on the  path $\theta (s)$
\beq
\cI U_{ab}[x,\theta ]~=~U_{ab}[x, \cI\theta]~.
\label{11a}
\eeq 
For linear combinations of $U$'s for different contours we extend $\cR $
and $\cI$ linearly.

Using the hermiticity of the matrices $A,~\phi$ in the form
$\overline{A}=A^t,~~\overline{\phi}=\phi ^t$ we can reformulate the
r.h.s in the second line of (\ref{11}) applying the following steps 
\bea
\overline{U_{ab}[x,\theta]}&=&\left (
P\exp\int _{s_i}^{s_f}\{-iA^t(x(s))\dot x (s)+\phi ^t(x(s))\theta (s)\vert \dot x\vert\}ds\right )_{ab}\nonumber\\
&=&\left (
\hat{P}\exp\int _{s_i}^{s_f}\{-iA(x(s))\dot x (s)+\phi (x(s))\theta (s)\vert \dot x\vert\}ds\right )_{ba}~.
\label{12}
\eea
Here $P,~\hat{P}$ denote ordering of matrices from right to left with
increasing/decreasing argument $s$. $\hat{P}$ applied to the path $x$ yields the
same result as $P$ applied to the backtracking path
\beq
(\cB x)(s)~=~x(s_f+s_i-s),~~~(\cB \theta )(s)~=~\theta (s_f+s_i-s)~.
\label{13}
\eeq
Therefore, we get
\beq
\overline{U_{ab}[x,\theta]}~=~U_{ba}[\cB x,\cB\theta]~.
\label{14}
\eeq
This, combined with (\ref{11}),(\ref{11a}) yields finally
\beq
\cR \cI U_{ab}[x,\theta ]~=~U_{ba}[\cB\cR x,\cB\cI\theta]~.
\label{15}
\eeq
It is worth pointing out that for the result (\ref{14}) the presence/absence of the
factor $i$ in front of the $A$ and $\phi $ term in $U$ is crucial. One could
consider this as another argument for the choice favoured by the investigations
of ref. \cite{gross}.\\

We now turn to a derivation of the basic Osterwalder-Schrader positivity
condition in a streamlined form within the continuum functional integral
formulation. All steps can be made rigorously by a translation into a lattice
version with local and nearest neighbour interactions.

Let denote $H_{\pm}=\{x^{\mu}\vert \pm x^1>0\},~~~H_{0}=\{x^{\mu}\vert 
x^1=0\}$. Then we consider for a functional of two paths 
$x^{(1)},x^{(2)}\in H_+$
\beq
f[x^{(1)},\theta ^{(1)};x^{(2)},\theta ^{(2)}]~=~U_{ab}[x^{(1)},\theta ^{(1)}]
~+~\lambda U_{ab}[x^{(2)},\theta ^{(2)}],~~~~\lambda ~~\mbox{real}~,
\label{15a}
\eeq
\bea
\langle f[x,\theta ]\cR \cI f[x,\theta ]\rangle&=&\int \cD A\cD \phi
f[x,\theta ]\overline{f[\cR x,\cI\theta ]}~e^{-S}
\\
&=&\int \cD A^{(0)}\cD \phi ^{(0)}~e^{-S_0}\nonumber\\
&\cdot &\int _{(b.c.)}\cD A^{(+)}\cD \phi ^{(+)} f[x,\theta ]~e^{-S_+}\cdot
\int _{(b.c.)}\cD A ^{(-)}\cD \phi ^{(-)}\overline{f[\cR x,\cI\theta ]}~e^{-S_-}~.
\nonumber
\label{15b}   
\eea
$\pm ,~0$ on the fields as well as on the action indicates that
it refers to points in $H_{\pm},~H_{0}$. The index for the two paths has been
dropped, and the boundary condition $(b.c.)$ is
$$A^{(\pm )}\vert _{\partial H_{\pm}}=A^{(0)},~~~\phi ^{(\pm )}\vert
_{\partial H_{\pm}}=\phi ^{(0)}~.$$ 
With the abbreviation
\beq
h[A^{(0)},\phi ^{(0)},x,\theta ]~=~\int _{(b.c.)}\cD A^{(+)}\cD \phi ^{(+)}
f[x,\theta ]~e^{-S_+}~,
\label{15c}
\eeq
the standard reflection properties of the action imply
\beq
\langle f[x,\theta ]\cR \cI f[x,\theta ]\rangle ~=~\int \cD A^{(0)}\cD \phi
^{(0)}~e^{-S_0}~h[A^{(0)},\phi ^{(0)},x,\theta ]\cdot \overline{h[A^{(0)},\phi
  ^{(0)},x,\cI\theta ] }~.
\label{15d}
\eeq
For $\cI = ${\bf 1}$ $ the integrand of the final integration over the fields
in the reflection hyperplane $H_0$ is non-negative, hence
\beq
\langle f[x,\theta ]\cR  f[x,\theta ]\rangle ~\geq ~0~.
\label{15e}
\eeq

For nontrivial $\cI $ the situation is by far more involved. If there would be
no boundary condition, the result of the half-space functional integral in
(\ref{15c}) would be invariant with respect to 
$\theta\rightarrow\cI\theta $. A given boundary configuration in general breaks
$O(6)$ invariance on $S^5$. But due to the $O(6)$ invariance of the action,
the functional integration measure and the $\phi\theta$ coupling in $f$, we have instead
\beq
h[A^{(0)},\cI\phi ^{(0)},x,\cI\theta ]~=~h[A^{(0)},\phi ^{(0)},x,\theta ]~.
\label{15g}
\eeq
This implies
\bea
\langle f[x,\theta ]\cR \cI f[x,\theta ]\rangle ~=~\int \cD A^{(0)}\cD \phi
^{(0)}~e^{-S_0}\label{15h} \\
\cdot~\frac{1}{2}\left (~h[A^{(0)},\phi ^{(0)},x,\theta ]~\overline{
h[A^{(0)},\phi ^{(0)},x,\cI\theta ] }\right .&+&\left . h[A^{(0)},\phi ^{(0)},
x,\cI ^{-1}\theta ]~\overline{h[A^{(0)},\phi ^{(0)},x,\theta ]}~\right ) ~,
\nonumber
\eea
which says us only ($R$ real numbers)
\beq
\langle f[x,\theta ]\cR \cI f[x,\theta ]\rangle \in R ~~~~\mbox{for}~~
\cI ^2={\bf 1}~.
\label{15i}
\eeq   
The statements (\ref{15e}) and (\ref{15i}) are rigorous ones. 
Beyond them we found no real proof for sharpening (\ref{15i}) to an inequality
of the type (\ref{15e}) for some nontrivial $\cI $. For later application
to the estimate of rectangular Wilson loops we are in particular interested
in nontrivial isometries keeping the, by assumption common, $S^5 $ position
of the endpoints of the contours on $H_0 $ fixed. Then $\cI =\cI_{\pi}$,
denoting a rotation around this fixpoint with angle $\pi $, are the only
candidates.

At least for boundary fields $\phi ^{(0)}$ in (\ref{15c}), which as a map
$R^3\rightarrow S^5$ have a homogeneous distribution of their image points on
$S^5$, we can expect that for contours of the type discussed in connection 
with fig.1 below in the limit of large $T$ the orientation of $\theta $
relative to $\phi ^{(0)}$ becomes unimportant. Therefore, we conjecture for this special situation  
\beq
\langle f[x,\theta ]\cR \cI_{\pi} f[x,\theta ]\rangle ~\geq ~0~.
\label{15j}
\eeq
From (\ref{15e}) and (\ref{15j}) for any real $\lambda $ in (\ref{15a})
we get via the 
standard derivation of Schwarz-type inequalities
\bea
\langle U_{ab}[x^{(1)},\theta ^{(1)} ]~\cR\cI U_{ab}[x^{(2)},\theta ^{(2)} ]
\rangle ^2&\leq &\langle U_{ab}[x^{(1)},\theta ^{(1)} ]~\cR\cI
U_{ab}[x^{(1)},\theta ^{(1)}]\rangle \label{15f}\\
&&~~~~~~~~~~~~\cdot\langle U_{ab}[x^{(2)},\theta ^{(2)} ]~\cR\cI
U_{ab}[x^{(2)},\theta ^{(2)}]\rangle ~.
\nonumber
\eea
This is a rigorous result for $\cI ={\bf 1}$ and a conjecture for $\cI =\cI _{\pi}$. 
\\ 

Let us continue with the discussion of a Wilson loop for a
closed contour which crosses the reflection hyperplane twice and which
is the result of going first along $x^-\in H_-$ and then along $x^+\in H_+$.
In addition we restrict to cases of coinciding $S^5$ position at the
intersection points with $H_0$ and treat in parallel $\cI ={\bf 1},\cI _{\pi}$
\bea
W[x^+\circ x^-,\theta ^+\circ \theta ^-]~=~\sum _{ab}
\langle U_{ab}[x^+,\theta ^+]~U_{ba}[x^-,\theta ^-]\rangle
~~~~~~~~~~~~~~~~~~~~~~~~~~~~\label{19}\\
=~\sum _{ab}\langle U_{ab}[x^+,\theta ^+]~\cR\cI\overline{U_{ba}[\cR x^-,\cI ^{-1}\theta ^-]}\rangle ~~~~~~~~~~~~~~~~~~~~~~~~~\nonumber\\
\leq \sum _{ab}\langle U_{ab}[x^+,\theta ^+]~\cR\cI U_{ab}[x^+,\theta^+]\rangle ^{\frac{1}{2}}\langle\overline{U_{ba}[\cR x^-,\cI ^{-1}\theta ^-]}~
\cR\cI \overline{U_{ba}[\cR x^-,\cI ^{-1}\theta ^-]}\rangle ^{\frac{1}{2}} 
\nonumber\\
\leq (
\sum _{ab}\langle U_{ab}[x^+,\theta ^+]~\cR\cI U_{ab}[x^+,\theta ^+]\rangle
)^{\frac{1}{2}}
~(\sum _{cd}\langle \overline{U_{cd}[\cR x^-,\cI ^{-1}\theta ^-]}~\cR\cI 
\overline{U_{cd}[\cR x^-,\cI ^{-1}\theta ^-]}\rangle
)^{\frac{1}{2}}.
\nonumber
\eea
We have used (\ref{11}), (\ref{11a}), $\cR\cR x=x$, (\ref{15f}) and the usual Schwarz inequality
in the last step. Now with (\ref{14}),(\ref{15}) we get
\bea
W[x^+\circ x^-,\theta ^+\circ \theta ^-]
&\leq &\left (
W[x^+\circ\cB\cR x^+,\theta ^+\circ \cB\cI\theta ^+]
\right )^{\frac{1}{2}}\nonumber\\ 
&&~~~~~~~\cdot ~\left (
W[\cB\cR x^-\circ x^-,\cB \cI ^{-1}\theta ^-\circ \theta ^-]\right )^
{\frac{1}{2}}~.
\label{20}
\eea 
\begin{figure}
\begin{center}
\mbox{\epsfig{file=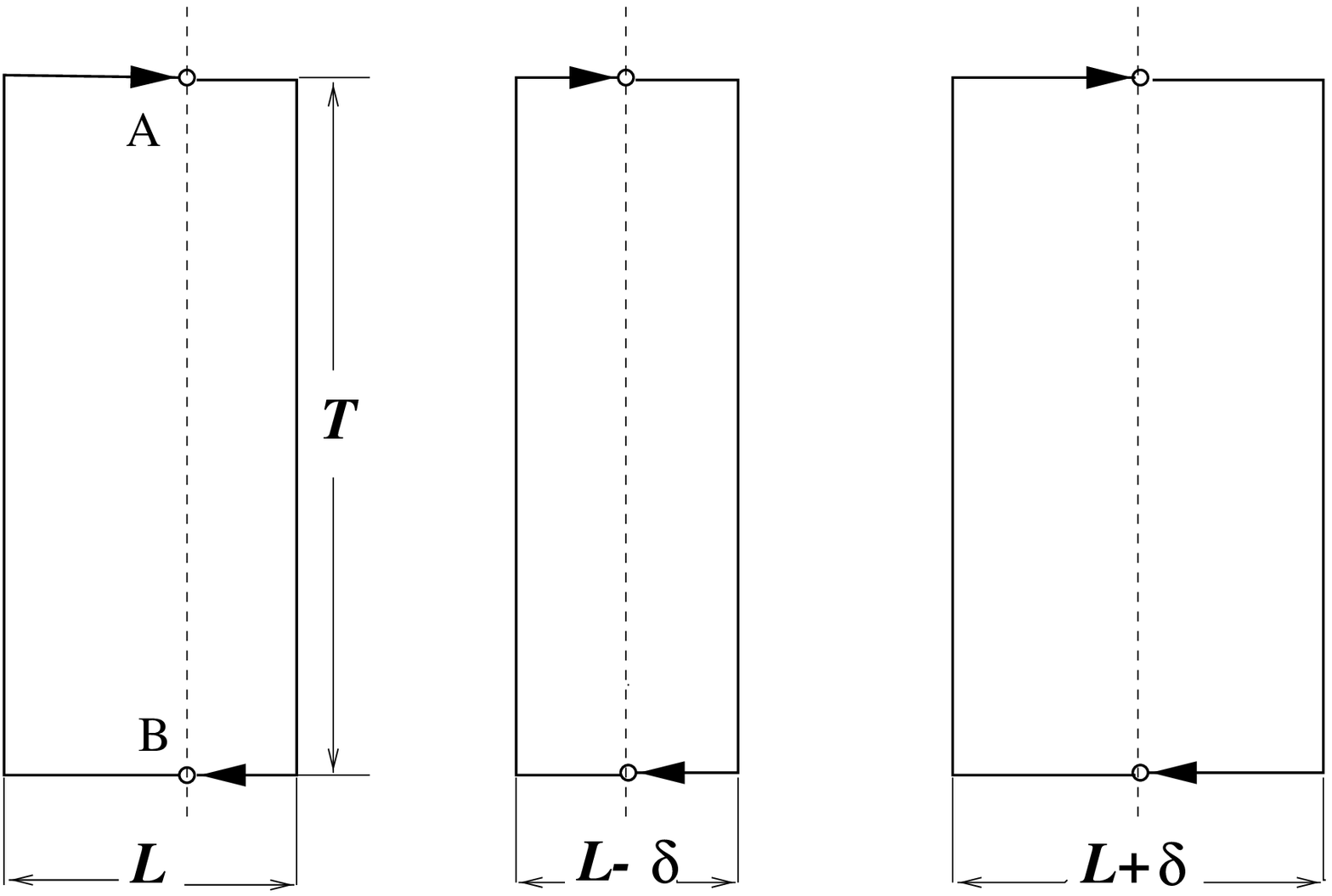, width=100mm}}
\end{center}
\noindent {\bf Fig.1}\ \ {\it From left to right the contours
$x^+\circ x^-,~~x^+\circ \cB\cR x^+,~~\cB\cR x^-\circ x^-$.}
\end{figure}

To evaluate the potential between two static sources ($Q\bar Q $) separated
by the distance $L$ and located at fixed $S^5 $-positions $\theta _Q,~\theta _
{\bar Q}$ we need Wilson loops for rectangular contours of extension $L\times T$ in the large $T$-limit. We choose the $S^5$-position on the two $L$-sides
linearly interpolating between $\theta _Q$ and $\theta _{\bar Q}$ on the corresponding great circle. For this restricted set of contours the Wilson loop becomes a  function of $L,~T$ and the angle between $\theta _Q$ and $\theta _{\bar Q}$, called $\Theta $. 

In addition it is useful to restrict ourselves to contours which are situated 
in planes orthogonal to the reflection hyperplane and with $T$-sides running
parallel to it in a distance $\frac{L\pm\delta}{2}$, see fig.1. Then 
$\cI =\cI_{\pi}$ reflects $\theta ^{\pm}(s)$, 
which both lie on the great circle through $\theta _Q$ and $\theta _{\bar Q}$,
with respect to the common $S^5$-position of the points $A$ and $B$, see 
fig.1. As a consequence, (\ref{20}) implies
\beq
W(L,T,\Theta)~\leq ~\left (W(L-\delta,T,\frac{L-\delta}{L}\Theta )\right )^{\frac{1}{2}}~\left (W(L+\delta,T,\frac{L+\delta}{L}\Theta )\right )^{\frac{1}{2}}~,
\label{21}
\eeq
which by standard reasoning yields for the static potential 
\beq
V(L,\Theta )~\geq ~\frac{1}{2}\left (V(L-\delta ,\frac{L-\delta}{L}\Theta )~+~
V(L+\delta ,\frac{L+\delta}{L}\Theta )\right )~.
\label{22}
\eeq
The last inequality implies the local statement 
$\frac{d^2}{d\delta ^2}V(L+\delta ,\frac{L+\delta }{L}\Theta )\leq 0$, i.e.
\beq
\left ( L^2~\frac{\partial ^2}{\partial L^2}~+~2L\Theta ~\frac{\partial ^2}{\partial L\partial \Theta }~+~\Theta ^2~\frac{\partial ^2}{\partial \Theta ^2}\right  )~V(L,\Theta )~\leq 0~.
\label{24}
\eeq
It means concavity on each straight line across the origin, in the relevant
part of the $(L,\Theta )$-plane, $0<L<\infty ,~~0<\Theta\leq\pi $.\\

Both (\ref{22}) and (\ref{24}) rely on the conjecture (\ref{15j}). 
From the rigorous point of view we are allowed to use (\ref{20}) 
for $\cI ={\bf 1}$ only. Then the paths generated on the r.h.s. are, with
respect to their $S^5$ properties, no
longer of the type with which we started on the l.h.s. On the part of the
space time contour orthogonal to $H_0$ we go e.g. from $\theta _{Q}$
to the common $S^5$ position of the points A and B and then $back$ to $\theta
_{Q}$. Since in the large $T$-limit, relevant for the extraction of the
$Q\bar Q $-potential, only the behaviour on the large $T$-sides matters,
we get
\beq
V(L,\Theta )~\geq ~\frac{1}{2}\left (V(L-\delta ,0 )~+~
V(L+\delta ,0 )\right )~.
\label{24a}
\eeq
This means standard concavity at $\Theta =0$ $and$
\beq
V(L,\Theta )~\geq ~V(L,0)~.
\label{24b}
\eeq  
If the same steps are repeated for rectangles with large $T$-sides still 
parallel to $H_0$, but spanning a plane no longer orthogonal to $H_0$
one finds
\beq
V(L,\Theta )~\geq ~\frac{1}{2}\left (V(\alpha (L-\delta ),0 )~+~
V(\alpha (L+\delta ) ,0 )\right )~,~~~0\leq \alpha\leq 1~.
\label{24c}
\eeq
The only new information gained from (\ref{24c}) is that $V(L,0)$ is
monotonically non-decreasing in $L$.
\section{Test of the generalised concavity condition for potentials derived via AdS/CFT duality}
The simplicity of the calculation recipe for Wilson loops in the classical
SUGRA approximation via AdS/CFT duality allows to make statements on universal
properties of the arising $Q\bar Q$-potential for a large class of SUGRA
backgrounds \cite{minahan, kinar}. We now enter a discussion of (\ref{24})
within this framework. The metric of the SUGRA background is assumed in the 
form
\beq
G_{MN}dx^Mdx^N~=~G_{00}(u)dx^0dx^0~+~G_{\vert\vert}(u)dx^mdx^m~+~G_{uu}(u)dudu~+~G_{\Omega}(u)d\Omega _5^2~.
\label{25}
\eeq
Then with
\beq
f(u)~=~G_{00}G_{\vert\vert },~~~g(u)~=~G_{00}G_{uu},~~~j(u)=G_{00}G_{\Omega }
\label{26}
\eeq
we get along the lines of \cite{malda, minahan, kinar, dorn}
\bea
L^{(\Lambda )}&=&2\sqrt{f_0}\sqrt{1-l^2}\int_{u_0}^{\Lambda}\sqrt{\frac{gj}{f}}
\frac{du}{\sqrt{j(f-f_0)+(jf_0-j_0f)l^2}}~,\nonumber\\
\Theta ^{(\Lambda )}&=&2l\sqrt{j_0}\int_{u_0}^{\Lambda}\sqrt{\frac{gf}{j}}
\frac{du}{\sqrt{j(f-f_0)+(jf_0-j_0f)l^2}}~,\nonumber\\
V^{(\Lambda )}&=& \frac{1}{\pi}\int_{u_0}^{\Lambda}\sqrt{gfj}~
\frac{du}{\sqrt{j(f-f_0)+(jf_0-j_0f)l^2}}~.
\label{27}
\eea
We defined $f_0=f(u_0)$ etc. $\Lambda $ is a cutoff at large values of $u$.
In the following our discussion will be restricted to values of $L$ and
$\Theta $ for which all expressions appearing under square roots above are
positive and where the inversion $u_0=u_0(L,\Theta ),~~l=l(L,\Theta )$ is well
defined. (\ref{27}) implies
\bea
V^{(\Lambda )}&=&\frac{1}{\pi}\int_{u_0}^{\Lambda}\sqrt{\frac{g}{fj}}\sqrt{j(f-f_0)+(jf_0-j_0f)l^2}\nonumber\\
&+&\frac{1}{2\pi}\sqrt{f_0}\sqrt{1-l^2}~L^{(\Lambda )}~+~\frac{1}{2\pi}
\sqrt{j_0}~l~\Theta ^{(\Lambda )}~.
\label{28}
\eea
Now we differentiate with respect to $u_0$ and $l$. After this $\Lambda $
can be sent to $\infty$ ending up with a relation for the renormalised
potential $V$:
\bea
\frac{\partial V}{\partial u_0}&=&\frac{1}{2\pi}\sqrt{f_0}\sqrt{1-l^2}~
\frac{\partial L}{\partial u_0}~+~ \frac{1}{2\pi}\sqrt{j_0}~l~
\frac{\partial \Theta}{\partial u_0}~,\nonumber\\
\frac{\partial V}{\partial l}&=&\frac{1}{2\pi}\sqrt{f_0}\sqrt{1-l^2}~
\frac{\partial L}{\partial l}~+~ \frac{1}{2\pi}\sqrt{j_0}~l~
\frac{\partial \Theta}{\partial l}~.
\label{29}
\eea
For $V$ defined by (\ref{27}) implicitly as a function of $L$ and $\Theta $
this means
\beq
\frac{\partial V}{\partial L}~=~\frac{1}{2\pi}\sqrt{f_0}\sqrt{1-l^2},~~~
\frac{\partial V}{\partial \Theta}~=~\frac{1}{2\pi}\sqrt{j_0}~l~,
\label{30}
\eeq
i.e. $V$ is monotonically nondecreasing both in $L$ and $\Theta $.
The monotony in $\Theta $ is in agreement with our rigorous result
(\ref{24b}).\\

Calculating now second derivatives one arrives at ($f'_0=\frac{df(u_0)}{du_0}$ etc.)
\bea
\left ( L^2~\frac{\partial ^2}{\partial L^2}~+~2L\Theta ~\frac{\partial ^2}{\partial L\partial \Theta }\right .&+&\left .\Theta ^2~\frac{\partial ^2}{\partial \Theta ^2}\right  )~V(L,\Theta )~=\label{31}\\
&=&\frac{1}{4\pi\sqrt{f_0j_0}}~(Lf'_0\sqrt{j_0(1-l^2)}+
\Theta j'_0\sqrt{f_0} ~l)~(L\frac{\partial u_0}{\partial L}+\Theta\frac{\partial u_0}
{\partial \Theta})\nonumber\\
&+&\frac{1}{2\pi\sqrt{1-l^2}}~(\Theta\sqrt{j_0(1-l^2)}
-L\sqrt{f_0}~l)~(L\frac{\partial l}{\partial L}+\Theta\frac{\partial l}
{\partial \Theta})~.
\nonumber
\eea
Neglecting for a moment the issue of internal space dependence by
restricting oneselves to the case $\Theta =l=0$, one finds usual concavity 
in $L$ from (\ref{31}) if $f'_0~\frac{\partial u_0}{\partial L}\leq 0$.
The last inequality is for $f'>0$ guaranteed by theorem 1 of ref.\cite{kinar}.
\footnote{Our $f$ and $g$ are called $f^2$ and $g^2$ in that paper.}

Therefore, for $\Theta =0$ standard concavity of $Q\bar Q$-potentials with respect to the 
distance in usual space is guaranteed for the wide class of SUGRA backgrounds
covered by theorem 1 of ref.\cite{kinar}.

However, due to the more complicated structure of the l.h.s. of (\ref{31})
for $\Theta\neq 0$ we did not found a similar general statement in the
generic case. We can only start checking (\ref{24}) case by case.\\

As our first example we consider the original calculation of Maldacena
\cite{malda} for the $AdS_5\times S^5$ background. The result was
($R^2=\sqrt{2g^2_{YM}N}$)
\beq
V(L,\Theta )~=~-~\frac{2R^2}{\pi}\frac{F(\Theta )}{L}~,
\label{32}
\eeq
with
\bea
F(\Theta )&=&(1-l^2)^{\frac{3}{2}}\left ( \int_1^{\infty}\frac{dy}{
y^2\sqrt{(y^2-1)(y^2+1-l^2)}}\right )^2~,\nonumber\\
\Theta &=&2l\int _1^{\infty}\frac{dy}{\sqrt{(y^2-1)(y^2+1-l^2)}}~.
\label{33}
\eea
Due to this special structure ($L\frac{\partial V}{\partial L}=-V,~L^2\frac
{\partial ^2}{\partial L^2}V=2V,~ \frac{\partial \Theta}{\partial u_0}=0$), (\ref{24}) is equivalent to
\beq
\Theta ^3~\frac{d^2}{d\Theta ^2}\left (\frac{F}{\Theta }\right )~\geq 0~.
\label{34}
\eeq
A numerical calculation of $\frac{F}{\Theta }$ confirms (\ref{34}) clearly,
see fig.2.\\

Next we discuss the large $L$ confining potential including internal
space dependence and $\ap $ corrections of the background derived in
\cite{dorn}. It has the form ($\gamma =\frac{1}{8}\zeta (3)R^{-6}$, $T$
temperature parameter)
\beq
V(L,\Theta )~=~\frac{\pi R^2T^2}{2}(1-\frac{265}{8}\gamma)\cdot L~+~
\frac{R^2}{4\pi}(1+\frac{15}{8}\gamma )~\frac{\Theta ^2}{L}~+~O(1/L^3)~.
\label{35}
\eeq
\begin{figure}
\begin{center}
\mbox{\epsfig{file=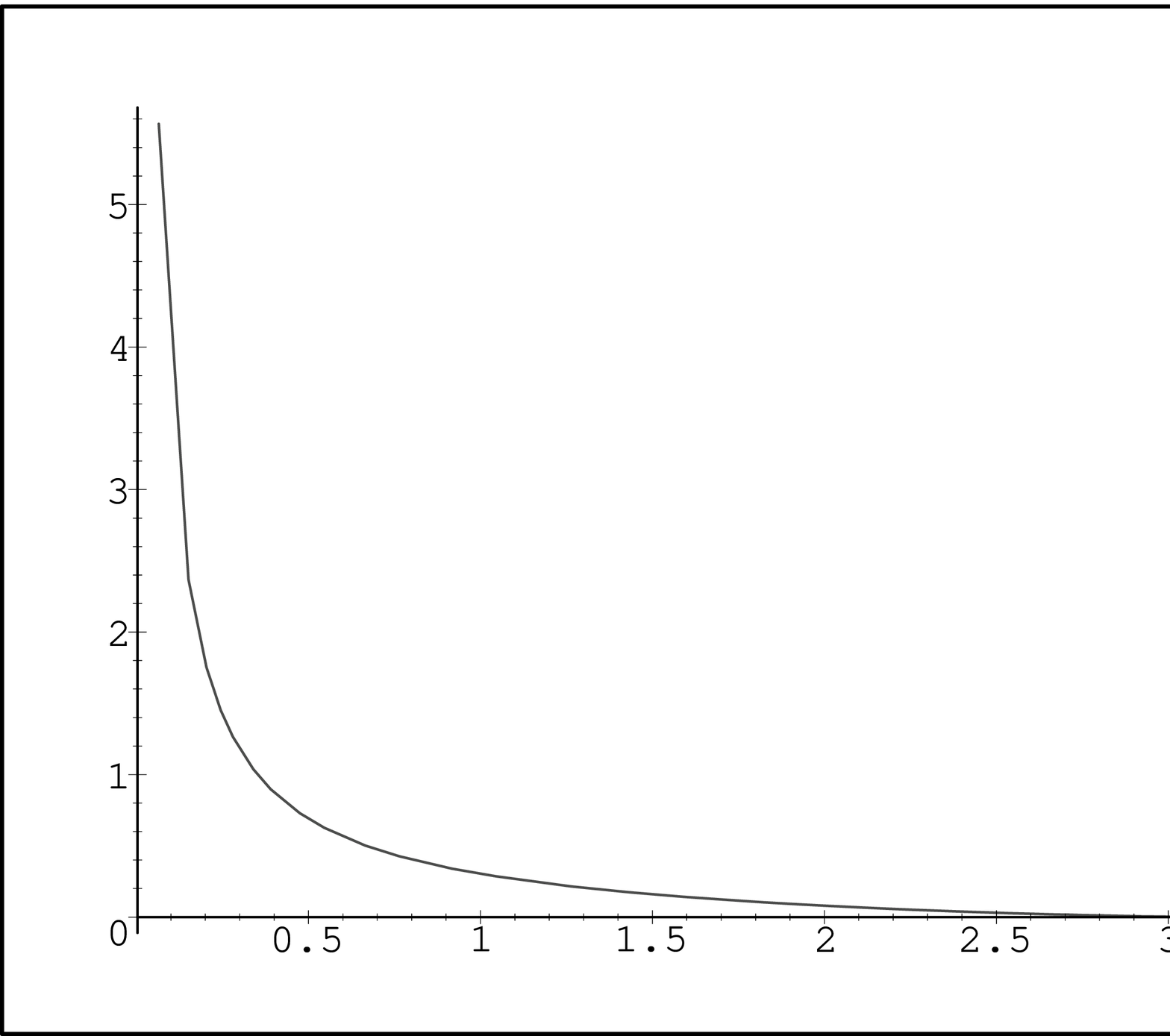, width=80mm}}
\end{center}
\noindent {\bf Fig.2}\ \ {\it $\frac{F}{\Theta}$ as a function of $\Theta$.
Use has been made of the representation in terms of elliptic integrals given
in \cite{malda}.}
\end{figure}
Although this potential for $\Theta \neq 0$ violates naive concavity
$\frac{\partial ^2V}{\partial L^2}\leq 0$, there is $no$ conflict with
the correctly generalised concavity (\ref{24}). Applied
to (\ref{35}) the differential operator just produces zero.
\section{Concluding remarks}
The $Q\bar Q$-potential derived \cite{malda} from the classical SUGRA approximation
for the type IIB string in $AdS_5\times S^5$  fulfils our generalised
concavity condition at $\Theta \geq 0$. This adds another consistency check
of this most studied case within the AdS/CFT duality. 

Potentials have been almost completely studied only for $\Theta =0$ in other
backgrounds. At least partly, this might
be due to the wisdom to approach in some way QCD, where after all there is no place for a parameter like this angle between different orientations in $S^5$.
However, one has to keep in mind that this goal, in the approaches discussed so far, requires some additional limiting procedure. Before the limit the full 10-dimensionality inherited by the string is still present. Fluctuation determinants
in all 10 directions have to be taken into account for quantum corrections
\cite{olesen,theisen} and the $\Theta $-dependence of the potentials is of 
course not switched off. 

Although we proved in classical SUGRA approximation monotony in $L$ and
$\Theta $ as well as concavity at $\Theta =0$
for a whole class of backgrounds, we were not able to get a similar general
result on concavity for $\Theta >0$. Further work is needed to decide, whether at all
general statements for $\Theta >0$ are possible. Alternatively one should
perform case by case studies for backgrounds derived e.g. from
rotating branes \cite{russo}, type zero strings \cite{tseytlin} or nonsupersymmetric
solutions of type IIB string theory \cite{sfetsos}.

On the field theory side further work is necessary to really prove
the conjectered inequality (\ref{15j}), otherwise the available set of
rigorous constraints on the $L$ $and$ $\Theta $ dependent potential,
beyond the standard concavity at $\Theta =0$, would contain only the very mild
condition (\ref{24b}).
\\[10mm]
{\bf Acknowledgement}\\
H.D. thanks G. Bali, H.-J. Otto and C. Preitschopf for
useful related discussions.
The work of V.D.P. was supported by GRACENAS grant, project No
97-6.2-34; RFBR-DFG grant, project No 96-02-00180 and RFBR grant,
project No 99-02-16617.

   
\end{document}